\documentstyle[graphicx]{aipproc}

\begin{document}

\title{Dielectric Response in Microscopically
Heterogeneous Dielectrics: Example of KTaO$_3$:Nb}

\author{S. Prosandeev$^{1}$,V. Trepakov$^{2,3}$, S. Kapphan$^{3}$,
M. Savinov$^{4}$}
\address{$^{1}$Rostov State University, 344090 Rostov on Don, 5 Zorge St.,  Russia;\\
$^2$A.F. Ioffe Physical \& Technical Institute, 194 021, St.
Petersburg, Russia;\\ $^{3}$FB Physik, University of Osnabruck,
49069 Osnabruck, Germany;\\ $^{4}$Instite of Physics AS CR, 182 21
Praha 8, Czech Republic}

\maketitle

\begin{abstract}

New experimental data on solid solutions of quantum paraelectrics
with KTaO$_3$:Nb as an example are considered within a framework
of a quantum theory of ferroelectric phase transitions. In order
to describe the effect of local heterogeneities a percolation type
theory together with a random field approach were employed.
\end{abstract}

\section{Introduction}

Solid solutions of quantum paraelectrics exhibit a variety of
intriguing properties, which have been attracting scientists for a
long period of time \cite{Huchli} but, in spite of this fact, some
important questions remain still under discussion. For example, it
is known that the substitution of Nb for Ta in KTaO$_3$ results in
the appearance of a strong temperature dielectric anomaly, which
is regarded to a ferrolectric phase transition, but at the
temperature of this anomaly and below manifestations of the
glass-type behaviour have been often reported. We consider these
effects in the framework of a random field approach combined with
a percolation theory that includes zero-point quantum vibrations.
New experimental data on the dielectric properties of the solid
solution KTaO$_3$:Nb (KTN) with $x=0.018$ (KTN 1.8) have been
obtained and discussed in connection with this problem.

\section{Quantum effects}

The quantum effect results in the following contribution to the
soft mode frequency $\Omega _c $ at zero wave vector \textbf{k}
\cite{Pros-Kleem}

\begin{equation}
\label{eq1}
\begin{array}{c}
 \Omega _c^2 = \omega _c^2 + d(T) \\
 \omega _c^2 = \omega _{c0}^2 + 3\beta P^2 + x\Xi \\
 d(T) = \frac{\displaystyle g_0 V_c }{\displaystyle 8\pi ^2}\int_{BZ} {d^3k\frac{\displaystyle 1}
 {\displaystyle \omega _{\rm {\bf
k}} }\left( {\coth \frac{\displaystyle \hbar \omega _{\rm {\bf k}}
}{\displaystyle 2k_B T} - 1} \right)}
\\
 \end{array}
\end{equation}

\noindent where $\omega _{c0} $ is the bare frequency at zero
temperature and zero wave vector, $P$ is polarization, $\Xi $ is
constant, $x$ is the impurity concentration, $\omega _{\rm {\bf
k}}^2 = \omega _c^2 + ck^2 + ...$, $g_0 $ is a constant
responsible for anharmonic interactions, $V_c $ is the unit cell
volume, and $k_B $ is the Boltzmann constant. Just below the phase
transition, one can neglect $\omega _c^2 $ in $\omega _{\rm {\bf
k}}^2 $, due to a large spatial dispersion of the soft mode in
KTaO$_{3}$, and, in this case, the integral in (1) is proportional
to $T^{2}$. Thus the temperature dependence of $\omega _c^2 $ at
$T_{c}$ is: $\omega _c^2 \sim T^2 - T_c^2 $, and at $T \ll \hbar
\omega _c $ $d(T)\sim T^{3 / 2}\exp \left( { - \hbar \omega _c^2 /
k_B T} \right)$. It is seen that the Currie-Weiss law is violated
not only in a small vicinity above $T_{c}$ as stressed in earlier
studies \cite{Rechester} but also below $T_{c}$.

Quantum effects can be observed in KTN 1.8 if one suppresses the
contribution connected with heterogeneities by applying a field
cooling procedure (see Fig 1). In this case a quadratic
temperature decrease of $\varepsilon (T)$ is obtained.
Polarization in quantum ferroelectrics just below $T_c$~ should
behave as  $P\sim \sqrt {T_c^2 - T^2} $ but at lower temperatures
polarization should saturate due to the zero-point quantum
vibrations \cite{Salje}. Thus the zero-point quantum vibrations
result in the saturation of the host-lattice polarization and
dielectric permittivity in ferroelectrics at low temperatures.

\begin{figure}[tbp]
\resizebox{0.8\textwidth}{!}{\includegraphics{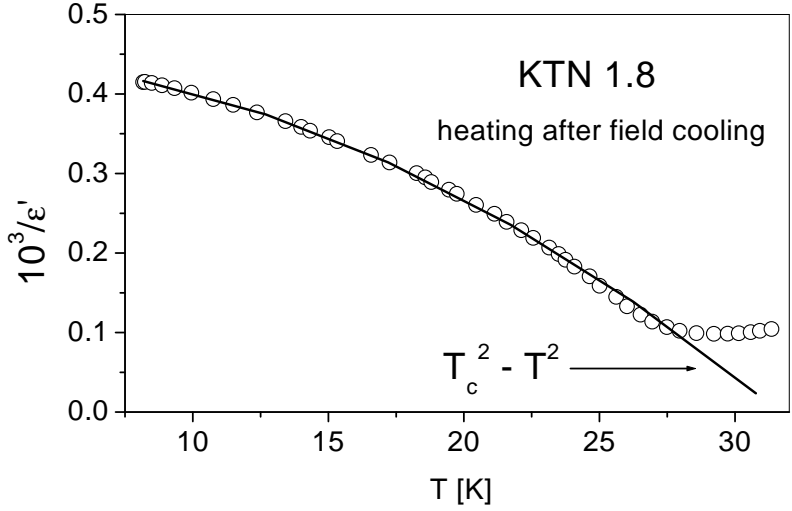}}
\caption{Inverse dielectric permittivity in KTN 1.8 obtained on
heating after field cooling} \label{Fig2}
\end{figure}

\section{Locally heterogeneous solid solutions}

\subsection{ Percolation approach}

The above analysis holds only for homogeneous ferroelectrics. In
our opinion, KTN 1.8 is not microscopically uniform due to
disorder in the impurity ion distribution over the corresponding
crystallographic positions \cite{Kleem1,Kleem2}. Indeed, since the
impurity concentration is very small, the average distance among
the impurities is rather large (e.g. in KTN 1.8 it is about 15.3 A
that is about 4 lattice parameters). In this case the fluctuations
of the average distance can be rather large also that leads to
percolative clustering of impurity ions and finally influences the
phase diagram of the corresponding solid solutions
\cite{Pros-Kleem,EPJ,Perc_IF}. Below we will discuss this behavior
in terms of the theory of percolation.

The impurity ions being close each other are correlated at
distances lower than $R_{cut - off} \sim \left( {V_0 + k_B T}
\right)^{ - 1 / 3}$ where $V_{0}$ is proportional to the sum of
the energies of the dipole-dipole and elastic impurity--impurity
interactions. At small impurity concentrations, which we consider
in the present study, the impurity clusters can appear only at low
temperatures where the cut-off radius is comparatively large. It
explains the experiment performed with KTN under high pressure
\cite{Samara}. When the pressure is normal the phase transition
occurs at large temperatures, at which the Nb clusters do not
appear, and, as a result, a frequency dispersion of the dielectric
permittivity is absent. When pressure is applied the soft mode
frequency increases and the phase transition temperature is
reduced to the region where the Nb clusters can already appear; as
a result, there appears a frequency dispersion of the
permittivity.

The average cluster size $<s>$ for interacting spheres with fixed
radii is usually given by a critical dependence \cite {Aharony}.
We consider the case when the interaction radius is temperature
dependent. In order to take this fact into account we introduce
the following dependence:

\begin{equation}
\label{<s>}
 < s > \sim \frac{1}{|(R_{cut - off}/a )^3x - c|^\gamma}
\end{equation}

\noindent where $\gamma $ is a critical exponent; $c$ is constant
and can be calculated on the basis of the percolation theory for
different lattices \cite{Aharony} (it equals 0.35 in the continual
percolation approach); the critical concentration can be found
from:

\begin{equation}
\label{x_c} x_c = \left( {a / R_{cut - off} } \right)^3c.
\end{equation}

It is seen that the critical concentration can be very small if
the cut-off radius is large enough. This finding corresponds to
experimental data for dilute solid solutions of quantum
paraelectrics according to which cluster phenomena appear in them
already at a very small impurity concentrations
\cite{EPJ,Pros-Kleem} that is a consequence of a large cut-off
radius. Note that the cut-off radius can increase not only due to
a straight dipole-dipole and quadrupole-quadrupole interactions
but also because of an indirect interaction over the soft mode.
The latter interaction can even diverge if the soft mode frequency
approaches zero but we do not consider this case here (see
\cite{EPJ}).

It follows from (\ref{x_c}) that when the cut-off radius increases
this can be considered as an increase of the unit volume (at
constant cut-off radius) and the effective concentration also
increases: $x_{eff} = (R_{cut - off} / a)^3x$. For us it is easier
to consider the percolation for spheres with the same (temperature
independent) radius but with the concentration of the spheres
effectively dependent on temperature instead of considering a
fixed impurity concentration but varying the sphere radius. In
this case a temperature decrease results in an \textit{ increase}
of the effective impurity concentration and due to this a
percolation phase transition can happen at some temperature.

After the substitution of the temperature dependent cut-off radius
value to (2) one can see that at the phase transition point the
average cluster size behaves as ${\left| T - T_{cp}
\right|^{-\gamma} }$(where $T_{cp} \sim x^{1 / 3} - const$ and $T
> T_{cp} )$. The most interesting result is that the critical
exponent $\gamma $ known in the theory of percolation as the
exponent for the critical \textit{concentration} dependence
coincides with the critical exponent for the critical
\textit{temperature} dependence. This exponent should be close to
$\gamma = 1.8$ as obtained in the theory of percolation that can
be a key to decide if the percolation approach is suitable or not
to describe concrete experimental data.

In reality expression (\ref{<s>}) is invalid for finite-size
heterogeneities. Unfortunately there is no analytical description
of such a situation although computations showed that the
singularity is diffused in this case \cite{Aharony}. In the
vicinity of the singularity one can try using the expression:

\begin{equation}
\label{s_diff}
 < s > \sim \frac{1}{|(R_{cut - off}/a )^3x - c|^\gamma+b^2}
\end{equation}

This expression differs from (\ref{<s>}) only in a vicinity of the
percolation threshold where the critical dependence is replaced by
a diffused anomaly.

\subsection{Heterogeneity size}

Similar to (\ref{s_diff}) the dielectric permittivity has rather a
rounding peak instead of a keen anomaly. It can be explained by
reaching the correlation radius the heterogeneity size. The
correlation radius, $r_{c}$, is the characteristic size of the
polarization fluctuation in the lattice. In uniform
ferroelectrics, when lowering temperature approaching $T_{c}$,
this radius increases as $r_c \sim \sqrt \varepsilon $. We
consider locally nonuniform ferroelectrics in the regime where the
main contribution to the dielectric permittivity stems from local
heterogeneities (for example in KTN these are regions enriched
with Nb) and where these heterogeneities can be considered, at
first glance, as independent. In this case the correlation radius
\textit{is not able to exceed } the heterogeneity size, which can
be defined as the maximal size of the polar regions . It implies
that the correlation radius will \textit{saturate} at lower
temperatures. Such a reason for this saturation differs from the
zero-point quantum vibrations: \textit{ it is connected with a
finite heterogeneity size}.

There are different possibilities to interpolate the temperature
dependence of the correlation radius described above from the
Curie-Weiss dependence at high temperatures to the constant
behavior at low temperatures. Earlier we used the Barrett formula
for this purpose \cite{Pros-Kleem,Li-pairs}.

It is important to notice that the considered saturation of the
correlation radius is a general property of locally heterogeneous
ferroelectrics: for example, we believe that it holds in relaxors
like PMN. Indeed, at temperatures above the Burns temperature a
Curie-Weiss behavior was evidenced but below the Burns temperature
and above the $T_m$ dielectric permittivity maximum temperature
rather a quadratic temperature dependence was observed. We have
found the following form of the dielectric permittivity suitable
in the whole temperature region \cite{Bokov}

\begin{equation}
\label{PMN}
 1/\chi = A(T-T_0)^2/T+f
\end{equation}

\noindent where $f$~ is constant. At temperatures close to $T_0$
this expression gives a quadratic temperature dependence, which
has zero derivative at $T=T_0$. At high temperatures this
expression gives the Curie-Weiss behavior. It is interesting to
notice that the same expression can be considered as a sum of a
linear temperature term, $A(T-2T_0)$, and an inverse temperature
term, $(AT_0^2+f)/T$.

Hence there is a temperature interval where the correlation radius
reaches the heterogeneity size and saturates thereafter (at lower
temperatures). In this interval there is a deviation from the
Curie-Weiss law due to the saturation of the correlation radius.
Additional (hydrodynamic) fluctuations appear at these
temperatures, which we consider in the next subsection.

\subsection{Orientable Polar Regions}

The main difference of the dielectric response in a nonuniform
dielectric media relative the uniform one, besides the appearance
of precursors described above in subsection A, is the existence of
\textit{added polarizability }due to the polar microregions. This
new feature of the nonuniform media manifests itself by a very
peculiar temperature dependence of the dielectric permittivity
below $T_{c}$. In Fig. 2 we plot the temperature dependence of the
inverse dielectric permittivity of KTN 1.8 obtained on heating. It
is seen that the low-temperature behavior is given by a straight
line, the inclination of which is noticeably lower than one for
the high-temperature branch but according to the theory of
ferroelectrics \cite{Levanjuk}, below $T_{c}$, it should be two
times larger than above $T_{c}$. This contradiction can be
explained if one takes into account the saturation of the
correlation radius at $T_c$ and a contribution of the polar
microregions to the dielectric permittivity below $T_{c}$. The
latter contribution can originate from ordering of the polar
region dipole moments in external field and due to the
interactions among the polar regions.

\begin{figure}[tbp]
\resizebox{0.8\textwidth}{!}{\includegraphics{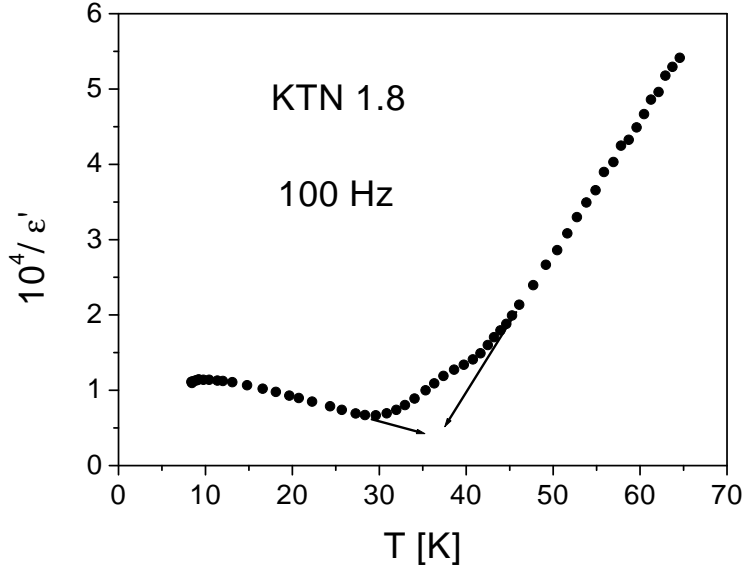}}
\caption{Inverse dielectric permittivity in KTN 1.8 below $T_c$}
\label{Fig3}
\end{figure}

Indeed, consider random fields \textbf{e} (see for definitions and
experimantal studies of the random fields Refs.
\cite{Kleem3,Vugmeister-Glinchuk}) and merged to their directions
local dipole moments ${\rm {\bf{ \mu} }}$. In the field ${\rm {\bf
E}} = {\rm {\bf E}}_{\rm {\bf 0}} + \eta {\rm {\bf P}}$, where
\textbf{E}$_{0}$ is the external field and \textbf{P} being the
polarization, the dipole moments are directed along \textbf{E} +
\textbf{e}. It results in the following polarization

\begin{equation}
\label{polariz}
\begin{array}{l}
 P = n\mu g(E) = \frac{\displaystyle n\mu }{\displaystyle 2}\int\limits_0^\pi {\sin \vartheta d\theta \left[
{\frac{\displaystyle E + e\cos \theta }{\displaystyle \sqrt {E^2 +
e^2 + 2Ee\cos \theta } } - \cos \theta } \right]} = \\
 \\
\,\,\,\,\,\,\,\,\,\,\,\,\,\,\,\,\,\,\,\,\,\,\,\,\,\,\,\, = \left\{
{{\begin{array}{*{20}c} {n\mu (1 - e^2 / 3E^2)} \hfill & {E > e}
\hfill \\ {2n\mu E / 3e} \hfill & {E < e} \hfill \\
\end{array} }} \right. \\
 \end{array}
\end{equation}

\begin{figure}[tbp]
\resizebox{0.8\textwidth}{!}{\includegraphics{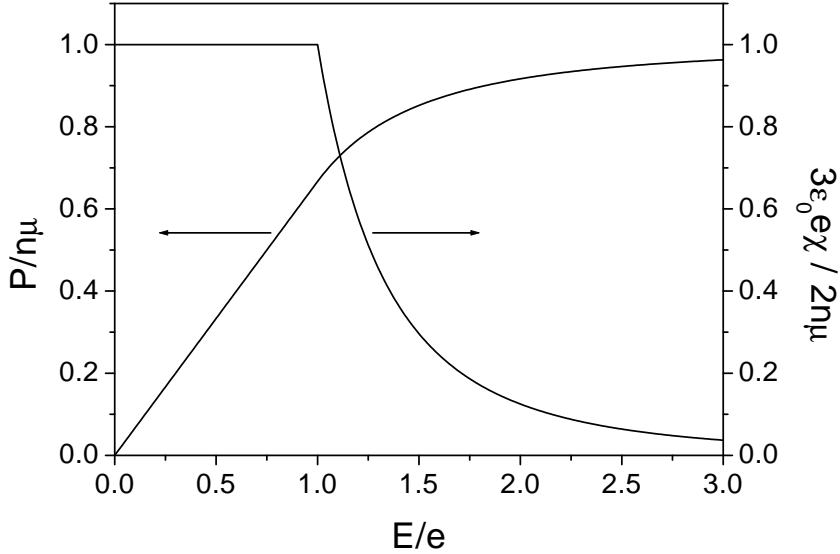}}
\caption{Model behaviour of the polarization and dielectric
susceptibility} \label{random}
\end{figure}

\noindent where $e=|\bf{e}|$. It follows from this result that the
susceptibility of nonineracting polar regions can be found from
the expression (see Fig. \ref{random}):

\begin{equation}
\label{suscept} \chi _0 = \left. {\frac{1}{\varepsilon _0
}\frac{dP}{dE}} \right|_{E = 0} = \left\{ {{\begin{array}{*{20}c}
 {2n\mu / 3e\varepsilon _0 } \hfill & {E < e} \hfill \\
 {2n\mu e^2 / 3\varepsilon _0 E^3} \hfill & {E > e} \hfill \\
\end{array} }} \right.
\end{equation}

\noindent where $n$ is the dipole (heterogeneity) concentration.
The temperature dependence of the dipole moment can be found from
the expression: $\mu = \mu _0 \tanh \left[ {\left( {{\rm {\bf \mu
}}_0 {\rm {\bf e}}} \right) / k_B T} \right]$ where ${\rm {\bf \mu
}}_0 $ is the dipole moment magnitude. This derivation explains
the existence of the large contribution to the dielectric
permittivity of KTN 1.8 below $T_{c}$ due to the Nb related
heterogeneities.

In the presence of macroscopic polarization the linear
susceptibility depends on the $P_{0}$ value, which can be found
from: $P_0 = \mu g(P_0 )$ and it appears at the condition $2n\mu
\eta / 3e = 1$, which provides $T_{c2} \sim \mu _0^2 n\eta / k_B
$. At large $E + \eta P_0 $ the contribution of the polar region
dipole moments to the dielectric permittivity rapidly decreases,
as $\left( {E + \eta P_0 } \right)^{ - 3}$. This explains the
absence of this contribution in our experimental data when the
sample was field cooled.

Obviously the macroscopic polarization obtained on field cooling
is larger than one got on zero field cooling. Hence the appearance
of macroscopic polarization suppresses the additional contribution
connected with the polar microregions. It implies that the new
phenomenon seen in the dielectric data of KTN 1.8 at low
temperatures can be explained by a very low value of macroscopic
polarization appeared on zero field cooling below $T_{c}$ because
of clustering the Nb impurities but without the appearance of the
connected cluster on the one hand and because of quantum effects
leading to suppression of the polarization growth at low
temperatures on the other hand. This peculiar situation results in
a random distribution of the local polarization over the polar
microregions merged to local random field directions and, as a
consequence, in an additional contribution to the dielectric
permittivity connected with the ordering of the polar region
dipole moments.

Our experimental data show a hysteresis phenomenon at low
temperatures, which can be explained within a Landau-type theory
if one assumes the existence of strong electrostriction
interaction in KTN 1.8 in agreement with experimental finding
\cite{Toulouse}. The final Hamiltonian can be written in the form

\begin{equation}
\label{eq6} H = \left( {\frac{2\mu _0 n}{3e}\coth \frac{\mu _0
e}{k_B T} - \eta } \right)\delta P^2 + \frac{1}{4}\tilde {\beta
}P^4 + \frac{1}{6}\xi P^6 + \upsilon \left( {\nabla P} \right)^2 -
EP
\end{equation}

\noindent where $\tilde {\beta } = \beta - 4\lambda / \kappa $,
$\lambda $ is the electrostriction constant and $\kappa $ being
the elastic constant. A large value of the electrostriction
constant leads to negative values of the nonlinearity constant
$\tilde {\beta }$ and, consequently, to the phase transition of
the first order.

To take into account the scattering of the random field magnitude
we used the following distribution function for a reorientable
part of the random fields \cite{Vugmeister-Glinchuk}

\begin{equation}
\label{distr} f(e) = \frac{1}{\left( {\sqrt \pi a} \right)^3}e^{ -
\left| {{\rm {\bf e}} - \eta {\rm {\bf P}}} \right|^2 / a^2}
\end{equation}

\noindent By integrating (\ref{suscept}) with this distribution
function we obtained at $E < e$

\begin{equation}
\label{eq8} \chi _0 = \frac{4n\mu }{3\varepsilon _0 \eta
P}erf\left( {\eta P / a} \right) \approx \frac{4n\mu }{3\sqrt \pi
\varepsilon _0 a}\left[ {1 - \frac{\eta ^2P^2}{3a^3} + ...}
\right]
\end{equation}

\noindent It is seen that the bare susceptibility (\ref{suscept})
decreases with the width of the distribution function
(\ref{distr}) and with polarization $P$.

The phase transition in the system consisting of polar regions can
be whether of the glass-type or ferroelectric. In order to decide,
which type of the phase transition will occur one can employ the
percolation technique again  but on the next, nanoscale level (see
also a consideration in the framework of a
Random-Field-Random-Bond model in \cite{Blinc}). One can introduce
a cut-off interaction radius for interactions among the polar
regions and obtain that if the polar region concentration exceeds
a critical concentration then a connected cluster appears and the
steady state is ferroelectric but in the reverse case only finite
clusters made of polar regions exist and the steady state is of
the glass nature.

Our experimental data show a strong frequency dependence of the
dielectric permittivity at $T_{c}$. We regard this finding to
potential barriers separating different positions of the random
fields. For example for Li impurities a six-well model can be
suitable to describe this situation \cite{six-well}. This model
considers dipoles embedded into elongated random fields having six
possible directions, which provides a description of a phase
transition when temperature or field are changed. In the electric
field this model gives a critical point. One can consider the
random fields coupled to the soft vibrations and these random
fields can be polarized by external field according to the
distribution function (\ref{distr}). The existence of the
potential barriers for the local random fields results in the
following frequency dependence of the dielectric permittivity
\cite{Pros-Kleem,Li-pairs}:

\begin{equation}
\label{eps-freq} \varepsilon ' \sim \frac{1}{\omega _{c}^2(T) -
\lambda^2nF(\omega)}
\end{equation}

\noindent where $F(\omega ) = \left[4k_BT\left(1 - i\left(\omega
\tau\right)^{1 - \alpha }\right)\right]^{-1}$. We introduced here
the Cole-Cole type frequency dependence bearing in mind a
relaxation time distribution. Here $\lambda $ is a coupling
constant. Expression (\ref{eps-freq}) at large temperature above
$T_{c}$ shows the Curie-Weiss behavior but at low temperatures
there is frequency dependent contribution, which is sufficiently
enlarged by the coupling of the local random fields with the soft
modes.

\begin{figure}[tbp]
\resizebox{0.8\textwidth}{!}{\includegraphics{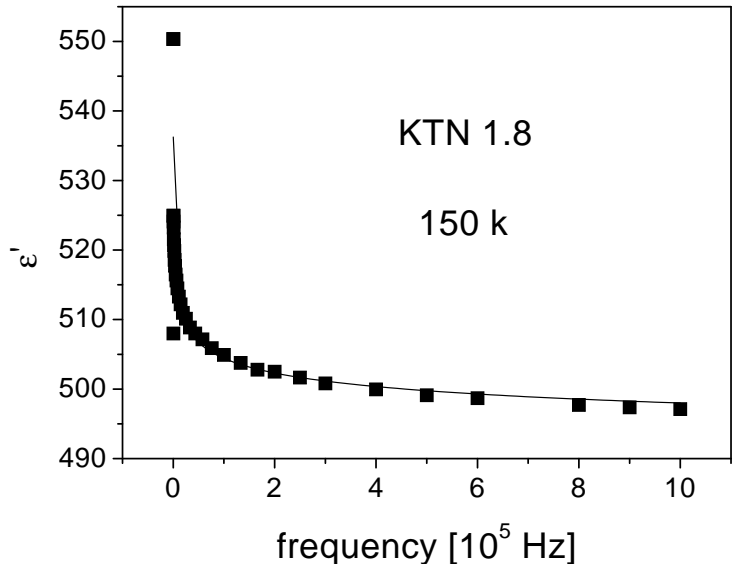}}
\caption{Universal relaxation in KTN 1.8 at high temperatures}
\label{Fig4}
\end{figure}

Besides the Cole-Cole contribution to the dielectric permittivity
dispersion we have found that even at large temperatures in KTN
1.8 there is universal dispersion of the form $\chi _U '(f)\sim
f^{n - 1}$ with the $n$~ value only slightly less than 1 (Fig. 4).
This frequency dependence can arise because of the existence of Nb
related heterogeneities or noncentrality even at large
temperatures. There can be also some influence of space charge on
this dispersion.

\section{Conclusions}

We showed that the striking dielectric properties of KTN 1.8 can
be understood and described on the basis of assumption that in
highly polarizable quantum paraelectrics even a small
concentration of dipolar impurity centres can form nano-scale
polar regions (clusters). The main addition to the dielectric
permittivity at high temperatures originates from these local
heterogeneities. At high temperatures, when the correlation radius
is larger than the heterogeneity radius, the heterogeneities
provide a Curie-Weiss law and an average over the bulk Curie
temperature governs the temperature dependence of the total
dielectric permittivity. When approaching the Curie temperature
the correlation radius increases but it saturates when reaching
the heterogeneity size. Together with quantum effects this leads
to the appearance of an intermediate state, in which the
dielectric permittivity is saturated. The orientation of the
cluster dipole moments as well as local random fields in an
external field results in the appearance of an additional
contribution to the dielectric permittivity in the low temperature
ferroelectric phase as it has been observed for KTN 1.8.

\section{Acknowledgements}

We thank NATO (PST. CLG.977348) 541, RFBR Grants 00-02-16875 and
01-02-16029. V.T. thanks DAAD for support.

\end{document}